\documentclass[superscriptaddress, pra, aps]{revtex4-2}

\newcommand{\ket}[1]{\left | #1 \right\rangle}
\newcommand{\bra}[1]{\left \langle#1 \right |}

\usepackage{epsfig}
\usepackage{xcolor}
\usepackage{amsmath}
\usepackage{amsfonts}

\begin{document}

\title[The fastest generation of multipartite entanglement with natural interactions]{The fastest generation of multipartite entanglement with natural interactions}

\author{Pawe{\l} Cie\'sli\'nski}

  \address{Institute of Theoretical Physics and Astrophysics, Faculty of Mathematics, Physics, and Informatics, University of Gdańsk, 80-308 Gdańsk, Poland}
  
  \author{Waldemar K{\l}obus}

  \address{Institute of Theoretical Physics and Astrophysics, Faculty of Mathematics, Physics, and Informatics, University of Gdańsk, 80-308 Gdańsk, Poland}
  
  \author{Paweł Kurzy\'nski}

  \address{Institute of Spintronics and Quantum Information, Faculty of Physics, Adam Mickiewicz University, 61-614 Pozna\'n, Poland}
  
  \author{Tomasz Paterek}

  \address{Institute of Theoretical Physics and Astrophysics, Faculty of Mathematics, Physics, and Informatics, University of Gdańsk, 80-308 Gdańsk, Poland}

\address{School of Mathematics and Physics, Xiamen University Malaysia, 43900 Sepang, Malaysia}
  
  \author{Wies{\l}aw Laskowski}
	
	\address{Institute of Theoretical Physics and Astrophysics, Faculty of Mathematics, Physics, and Informatics, University of Gdańsk, 80-308 Gdańsk, Poland}
		
	\address{International Centre for Theory of Quantum Technologies, University of Gdańsk, 80-308 Gdańsk, Poland}

\begin{abstract}
Natural interactions among multiple quantum objects are fundamentally composed of two-body terms only. In contradistinction, single global unitaries that generate highly entangled states {\color{black} typically} arise from Hamiltonians that couple multiple individual subsystems simultaneously. 
{\color{black}Here, we study the time to produce strongly nonclassical multipartite correlations with a single unitary generated by the natural interactions. We restrict the symmetry of two-body interactions to match the symmetry of the target states and} focus on the fastest generation of multipartite entangled Greenberger-Horne-Zeilinger (GHZ), W, Dicke  and absolutely maximally entangled (AME) states for up to seven qubits. These results are obtained by constraining the energy in the system and accordingly can be seen as state-dependent quantum speed limits for {\color{black}symmetry adjusted} natural interactions. 
They give rise to a counter-intuitive effect where the creation of particular entangled states with an increasing number of particles does not require more time. The methods used rely on extensive numerical simulations and analytical estimations.
\end{abstract}

\maketitle

\section{Introduction}
\label{sec:intro}

Entanglement is one of the central aspects of quantum mechanics. Its importance lies in the potential to revolutionise computing and communication as known today. Useful strongly entangled states are difficult to create and are usually obtained through a variety of at most two-qubit quantum gates \cite{Horodecki, Mora, Reck}. {\color{black} However, using sequences of noisy gates decreases the fidelity with the target state exponentially with the depth of the circuit.} One could solve this {\color{black} issue} by preparing certain useful states with a \emph{single} unitary acting on $N$ qubits at the same time. Many-qubit gates have indeed been studied in various contexts including, e.g. quantum circuits construction \cite{manyqubitgates1}, entanglement generation \cite{manyqubitgates2, manyqubitgates22, manyqubitgates222} and error correction \cite{manyqubitgates3, manyqubitgates4}.
{\color{black} In a simple example, a three-qubit Greenberger-Horne-Zeilinger (GHZ) state could be created from $\ket{000}$ by applying a unitary $U=|GHZ\rangle \langle 000| + \cdots$.} Unfortunately, this approach can lead to non-zero $3$-body  interaction terms in the Hamiltonian generating the evolution, i.e. $\mathrm{Tr} [ i\hbar/t\mathrm{log}(\mathrm{U}) \sigma_i \otimes \sigma_j \otimes \sigma_k]\neq 0$, where $\sigma_j$ ($j=x,y,z$) is the Pauli matrix.
While such interactions could be engineered in some systems~\cite{simulation1,simulation2,simulation3}, fundamentally physical couplings are pairwise.
The question, therefore, arises whether quantum states with multipartite {\color{black}nonclassical} correlations can be created through a single unitary evolution given by Hamiltonians with at most two-body terms. And further, how does this physically motivated constraint limit the speed of evolution? 

We answer these questions here and structure the paper as follows. In Sec.~\ref{sec:general} we define the physical setting and symmetries of the problem.
Next, in Sec.~\ref{sec:time-energy} quantum speed limit and normalisation of energy in the system are introduced. 
In Sec.~\ref{sec:3_qubits_estimation} we propose an intuitive estimation of energy standard deviation and use it to establish the minimal time for the creation of three-qubit GHZ{\color{black}} and W states via two-body interactions. Sec.~\ref{sec:numerics} presents numerical results for the optimisation of fidelity between the target and evolved state that provides the minimal time and exact analytical form of the generating two-body Hamiltonians for $N=3,...,7$ qubits in GHZ and W states. We also apply these methods to produce absolutely maximally entangled state (AME) of 5 qubits and give the minimal time for which the fidelity maximisation exceeds $99\%$.

\section{Physical setting}
\label{sec:general}

\begin{figure}[ht]
\centering
\includegraphics[width=0.65 \linewidth]{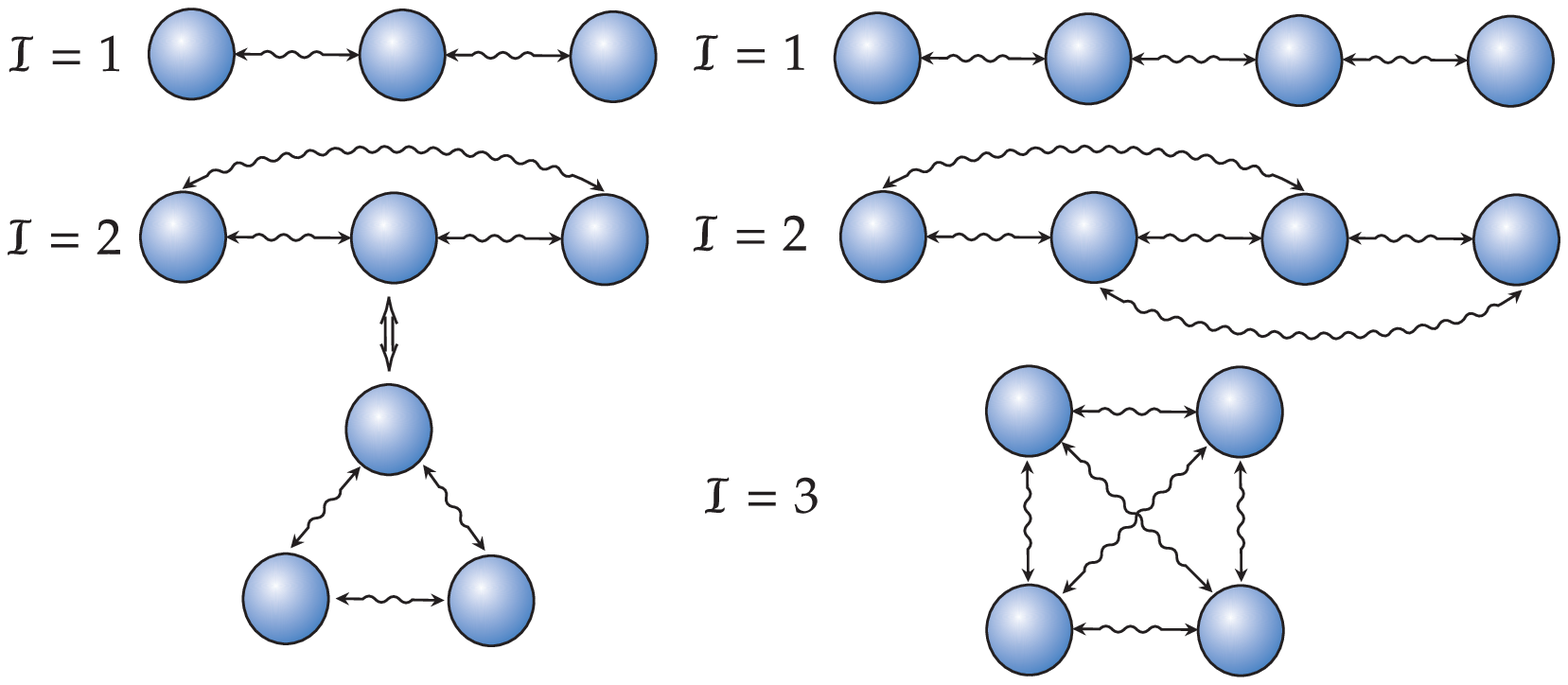}
\caption{\label{interaction} Interaction graphs. Various scenarios for interaction range in the case of $3$ and $4$ qubits studied in the article. Each edge represents two-body interaction between the particles. $\mathcal{I} = 1$ corresponds to nearest-neighbour coupling, etc.}
\end{figure}

Consider a general time-independent Hamiltonian with at most two-body interaction terms
\begin{equation}
    H=\sum_{i,j}\sum_{\mu_i,\mu_j} h_{\mu_i \mu_j}\sigma^{(i)}_{\mu_i}\sigma^{(j)}_{\mu_j}+b_{\mu_i}\sigma^{(i)}_{\mu_i},
\end{equation}
where the first sum includes the two-body interactions with strengths $h_{\mu_i \mu_j}$, 
$\sigma^{(i)}_{\mu}$ stand for Pauli matrices corresponding to the $i$th site ($\mu=x,y,z$), and $b_{\mu_i}$ is the strength of the on-site external field. 
The sums are over the particles coupled according to the interaction graphs presented in Fig.~\ref{interaction}.
We distinguish different ranges of interactions, $\mathcal{I}$. 
Interactions of range $\mathcal{I} = 1$ couple only nearest neighbours, of range $\mathcal{I} = 2$ couple more distant particles, or correspond to a more complex topology. The interactions of the highest rank $\mathcal{I}=N-1$ are represented by the {\color{black} complete interaction graph}. This is a very broad class of models. Hamiltonians that are diagonal in the Pauli decomposition can be found, e.g. in the Heisenberg-like models \cite{Heisenberg} and the charge or flux superconducting qubit models \cite{superconductingqubits}. Non-diagonal symmetric elements can exist in the transverse exchange \cite{transverseexchange} as well as in the Kaplan-Shekhtman-Entin-Wohlman-Aharony interaction \cite{KSEA}.

Throughout most of the paper, we consider Hamiltonians that are isotropic and symmetric (full permutation symmetry), i.e. 
\begin{equation}
h_{\mu_i \nu_j}=h_{\mu_k \nu_l}=h_{\nu_k \mu_l}
\quad
\textrm{ and } \quad
b_{\mu_i}=b_{\mu_j}.
\end{equation}
We will provide time bounds on entangled state generation that are optimal within this broad family {\color{black} and note that this symmetry matches the symmetry of the target states.}

Our focus is on the GHZ, W, Dicke and AME states for up to seven qubits, but it is clear that the methods can be used for arbitrary states.{\color{black} In this study only the five-qubit AME state was considered explicitly due to the fact that for $N=3$ it coincides with the GHZ state and in the case of $N=4$ as well as $N=7$ it was proven that such states do not exist~\cite{Scott_ame2004,Huber_ame2017,Huber_ame2018}. 
The states under consideration are therefore written as follows:
\begin{gather}
    |GHZ \rangle =\frac{1}{\sqrt{2}}(|0\cdots 0\rangle +|1\cdots 1\rangle),
    \label{ghz}\\
    |W \rangle =\frac{1}{\sqrt{N}}(|10\cdots 0 \rangle +|01\cdots 0\rangle +\cdots+|00\cdots1 \rangle ),
    \label{w}\\
    |D_N^k\rangle = \binom{N}{k}^{-1/2} \sum |s_1 \cdots s_N \rangle, \quad s_i \in \lbrace 0,1\rbrace,
    \label{dicke}
\end{gather}
where the last sum is over all permutations of placing `1' on $k$ positions. The AME$(5,2)$ state examined in our work is given as 
\begin{align}
     \ket{\mathrm{AME}(5,2)} = \frac{1}{\sqrt{8}} (\ket{01111} + \ket{10011} + \ket{10101} + \ket{11100}  \nonumber \\  
     - \ket{00000}- \ket{00110} - \ket{01001} - \ket{11010}),
     \label{ame52}
\end{align}
and was first constructed in~\cite{AME5}.}
{\color{black}It should be noted that we have chosen the common initial state $\ket{0\cdots 0}$ in all of the analysed cases, regardless of different fidelities with the target states.}

Since we limit ourselves to the Hamiltonians with a certain symmetry, it is only possible to reach the states with the same invariance property. This fact can be easily seen using the symmetry operators $\mathcal{P}$, acting on pairs of particles, that leave $H$ invariant, i.e. $\mathcal{P}H\mathcal{P}^{-1}=H$. If state $\ket{\phi}$ is invariant under $\mathcal{P}$ then $U\ket{\phi}=| \phi' \rangle$ has the same symmetry (the symmetry is preserved at all times). This comes from the fact that if $H$ is invariant under $\mathcal{P}$ then by the series expansion $U$ has the same property. Then clearly $[U,\mathcal{P}]=\mathcal{P}U-U\mathcal{P}=0$ and
$\mathcal{P} | \phi' \rangle = {\color{black}\mathcal{P}} U \ket{\phi} = U {\color{black}\mathcal{P}} \ket{\phi} = U \ket{\phi} = | \phi'\rangle$. The Hamiltonian corresponding to the {\color{black} complete interaction} graph is invariant under all pairwise swaps of particles and imposes the restriction on unitarily available states to permutationally symmetric ones. Notice that such states are very common in quantum information \cite{rydbergcomputing, ionscomputing, dicke, telecloning, cloning}. 

\section{Time-energy trade off -- speed of evolution}
\label{sec:time-energy}

Since $H$ and $t$ are linear in the exponent of the unitary evolution, one can always shorten or extend its duration by scaling the energy. It is the structure of energy levels, i.e. the variance or mean energy, that describes the ``speed'' of evolution{\color{black}~\cite{QSLMT,QSLML, Deffner_2017}}. The result of Mandelstam and Tamm allows the calculation of {\color{black}the} quantum speed limit (QSL) for a given initial and final state based on their overlap {\color{black}\cite{QSLMT}}
\begin{equation}
    t\geq t_{min}=\arccos(|\langle \psi(0)|\psi(t)\rangle|) \frac{\hbar}{\color{black}{\Delta H}},
    \label{MT}
\end{equation}
where ${\color{black}\Delta H}$ is the standard deviation of the energy operator in the initial state. The natural question arises, is the lower bound on this QSL tight for Hamiltonians with just one- and two-body terms? To study this problem we normalise the Hamiltonian in the following way 
\begin{equation}
    H\rightarrow \frac{H-\openone E_{\min}}{\tilde{E}_{\max}},
\end{equation}
where $E_{\min}$ is the minimal eigenvalue of $H$ and $\tilde{E}_{\max}=E_{\max}-E_{\min}$ is the maximal eigenvalue of $H- \openone E_{\min}$. Note that this constraint does not fix the variance but the energy range available in the system. {\color{black}Thus, it can be seen as an energy bandwidth normalisation.} A similar approach was recently pursued by Ness \emph{et al.} in \cite{energyboundQSL}. Normalisation endows us with operators $H$ with eigenvalues $E_i \in [0,1]$ in the natural units, which will be used throughout the rest of this paper. 

\section{Two-body interaction speed limits for three qubits}
\label{sec:3_qubits_estimation}

The key to our investigation of the effects of at most two-body interactions is hidden in the standard deviation of $H$. Indeed, the general speed limit (\ref{MT}) shows that in order to minimise the time required to produce a particular state, we should maximise the energy standard deviation. Intuitively, this quantity is maximised by the degenerate Hamiltonians because the energy levels can differ most in such cases and the eigenstates can be chosen such that the initial $\ket{0\cdots 0}$ state admits the mean energy right in the middle between the energy levels.

This path is followed rigorously in~\ref{sec:appI.1} where we show that the Hamiltonian generating the W state can be chosen without any {\color{black} on-site} terms and admits only two eigenvalues. 
According to our normalisation, the two energy levels are $E_0 = 0$ and $E_1 = 1$.
This immediately leads to $\Delta H = \sqrt{p_1(1-p_1)}$, where $p_1$ is the probability that the initial state has energy $E_1$. The maximal energy standard deviation is therefore $\Delta H = 1/2$ and the corresponding minimal time is given by
\begin{equation}
    t_{\min}^{\mathrm{W},2} = \frac{\pi}{2}\frac{1}{1/2 }= \pi,
\end{equation}
where $2$ in the superscript refers to {\color{black}isotropic and symmetric} two-body Hamiltonians. 

According to our analysis, there is no Hamiltonian with only two eigenvalues, {\color{black}within the considered class}, that produces the GHZ state{\color{black}, see~\ref{sec:appI.2}}. The required energy spectrum has to involve at least five eigenvalues and the {\color{black} on-site} terms turn out to be necessary. An exemplary Hamiltonian achieving the goal admits $\Delta H = 1/8$. This translates to
\begin{equation}
    t_{\min}^{\mathrm{GHZ},2} = \frac{\pi}{4}\frac{1}{1/8}= 2 \pi.
\end{equation}
This is similar to the geometrical approach in which the geodesics on two-dimensional subspace yield the fastest evolution \cite{Brody_2006, Aharonov} while engaging more eigenstates leads to a slowdown \cite{Bender2009}. Both of these times match our numerical results presented in Sec.~\ref{sec:numerics}. {\color{black} An explicit form of the generating Hamiltonians is provided in~\ref{sec:appII}.} 
But before we get there let us compare these new two-body speed limits with the three-body ones emerging from Eq.~(\ref{MT}).

\subsection{Comparison with three-body interaction speed limit}

A special case of three-body Hamiltonians are operators of the form $H=\sum_i \tilde{h}_i \sigma_i \otimes \sigma_i \otimes \sigma_i$, where $\tilde{h}_i$ give the strength of three-body interaction. Their spectrum consists of the minimal number of possible eigenvalues
\begin{equation}
    E=\pm \sqrt{\tilde{h}_x^2+\tilde{h}_y^2+\tilde{h}_z^2},
\end{equation}
and therefore they lead to the evolution times
\begin{equation}
    t_{\min}^{\mathrm{W},3} = \frac{\pi}{2}\frac{1}{1/2 }= \pi,
\end{equation}
\begin{equation}
    t_{\min}^{\mathrm{GHZ},3} = \frac{\pi}{4}\frac{1}{1/2}=  \frac{\pi}{2},
\end{equation}
which saturate the QSL. 
This is hardly a surprise since the optimal Hamiltonian contains $N$-body terms in general~\cite{Brody_2006}. The GHZ Hamiltonian is purely a three-body operator, whereas the $W$ Hamiltonian is the same as in the two-body interaction case. 
Accordingly, there is no speed increment due to tripartite coupling for the W state, but the GHZ state is produced four times faster.

The above predictions are consistent with the numerical calculations presented in the next section. For $N>3$ we will proceed with numerics only due to the order of the characteristic polynomials found in higher dimensions. Note that even for $N=3$ the polynomial is in general of the 8th order. The above estimations were possible due to the symmetry of $H$ and focus on the number of its distinct eigenvalues.

\section{Fastest entanglement generation for multipartite systems}
\label{sec:numerics}

We study unitary evolutions generated by Hamiltonians with at most two-body interaction terms through numerical computations. In order to answer our questions we maximise the fidelity between the target state $\ket{\psi}$ and the state produced at time $t$ by different Hamiltonians:
\begin{equation} 
  \mathcal{F}(t)=\underset{h_{\mu \nu},b_{\mu}}{\mathrm{max}} |\bra{\psi}\exp(-i H t)\ket{0 \cdots 0}|^2.
\end{equation}
This procedure yields $t_{\min}$ for each target state and allows us to look into the ``fastest'' Hamiltonians more closely. 
For the numerical optimisation, we performed parallel calculations using Wolfram Mathematica 13.0 and a random search-like algorithm {\color{black}written in Python}. {\color{black} In both environments time variable was discretised into steps $\Delta t$ ranging between $10^{-4}$, when the fidelity grows rapidly, to $10^{-1}$ where higher temporal precision is not required. For each time $t_k$ numerical maximisation of $|\langle \psi| \exp(-iH t_k)|0\cdots0 \rangle |^2$ is performed, with Hamiltonian re-normalisation after each choice of the parameters. In Mathematica, we used build-in numerical optimisation methods that included an evolutionary computation method called Differential Evolution and Random Search with $10^4$ iterations. For the Python code, we sampled $100-1000$ random initial parameters from a flat distribution and used the maximisation SciPy BFGS algorithm \cite{scipy} to find the local maximum of fidelity for each sampling. Other optimisers from SciPy as well as genetic algorithms from PyGad library~\cite{Pygad} were studied and the aforementioned one exhibited the best performance.} All data sets {\color{black}acquired from both environments} were then compared in order to find the maximal fidelity.
According to the applied numerical methods, we consider fidelity approximation up to the sixth decimal place. 
Each point on the obtained plots {\color{black} in Fig.~\ref{three_qubits}, Fig.~\ref{four_qubits} and Fig.~\ref{speedlimits}}, represents a different Hamiltonian that is optimal for a given time $t$. The final results on the two-body quantum speed limits for the GHZ, W and AME$(5,2)$  {\color{black}states obtained in this section} are gathered in Table.~\ref{res}. {\color{black} For comparison, we also present there the Mandelstam-Tamm quantum speed limits.} 

\begin{table}[]
\centering
\begin{tabular}{|c|c|c|c|c|c|c|}
\hline
 state & $N=3$ & $N=4$ & $N=5$  & $N=6$ & $N=7$ & {\color{black}MT} \\
 \hline
 GHZ &  $ 2 \pi \approx 6.28 $ & $2 \pi \approx 6.28 $ & $\frac{9\pi}{2}  \approx 14.14$  & $\frac{9\pi}{2} \approx 14.14 $ & $8 \pi \approx 25.13$ &$\frac{\pi}{2}  \approx 1.57$ \\
 W  & $ \pi \, \approx 3.14 $  & $\frac{\sqrt{11}\pi}{\sqrt{2}} \approx 7.37$ & $\frac{9\pi}{\sqrt{5}} \approx 12.64 $  & $\approx 18.76$ & $\approx 25.60$ & $ \pi \approx 3.14$ \\
 AME & $2\pi  \approx 6.28 $  & --- & $\approx 10.72^*$  &  & ---  & $\approx 2.42$ \\
 \hline
\end{tabular}
\caption{Summary of two-body quantum speed limits for {\color{black} complete interaction graphs obtained in Sec.~\ref{sec:numerics}}. All the times except for the AME state (marked with a star) refer to fidelity equal to 1, up to the sixth decimal place.
For the AME state we report the time corresponding to fidelity $0.99$ due to an extremely long plateau afterwards, see the discussion is Sec.~\ref{sec:ame}. {\color{black} For comparison, the last column gives the Mandelstam-Tamm (MT) quantum speed limits. This bound is independent of the number of particles for the GHZ and W states. In the case of AME states, the MT quantum speed limit was given for $N=5$.   }}
\label{res}
\end{table}

\subsection{Three and four qubits}

\begin{figure}[ht]
\centering
\includegraphics[width=0.57\linewidth]{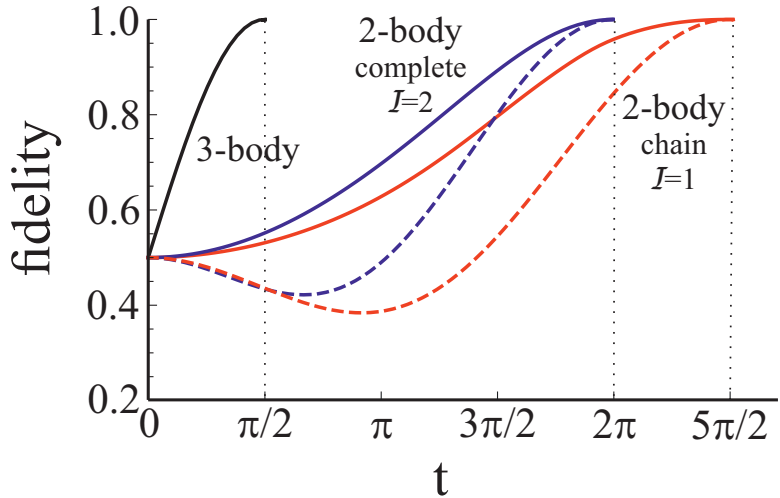}
\caption{\label{three_qubits} The fastest production of three-qubit GHZ state. The plot shows maximal fidelity $\mathcal{F}$ to the three-qubit GHZ state. 
{\color{black} The solid black line is for three-body interactions, the solid blue line for two-body interactions described by 
the complete interaction graph ($\mathcal{I}=2$) and the solid red line is for the linear chain ($\mathcal{I}=1$) scenario.} All solid lines were obtained by optimising over Hamiltonians at every time instance. Exemplary optimal evolutions generated by a single Hamiltonian are plotted in dashed lines.}
\end{figure}

The two-body GHZ quantum speed limits resulting from the optimisation are depicted in Fig.~\ref{three_qubits} and are consistent with the results obtained in the previous section. The three-body Hamiltonians for the GHZ state are, as expected, optimal and saturate the QSL. The {\color{black} complete interaction graph} is leading to the smallest $t_{\min}$ in the two-body regime as it contains more interaction terms. The linear chain interaction graph ($\mathcal{I}=1$) Hamiltonian performance is slower by the factor of $4/5$ ($t_{\min} = 5 \pi/2$). Based on this we proceed with the {\color{black} complete interaction graph} setting only in the next subsections. 

\begin{figure}[ht]
\centering
\includegraphics[width=0.9\linewidth]{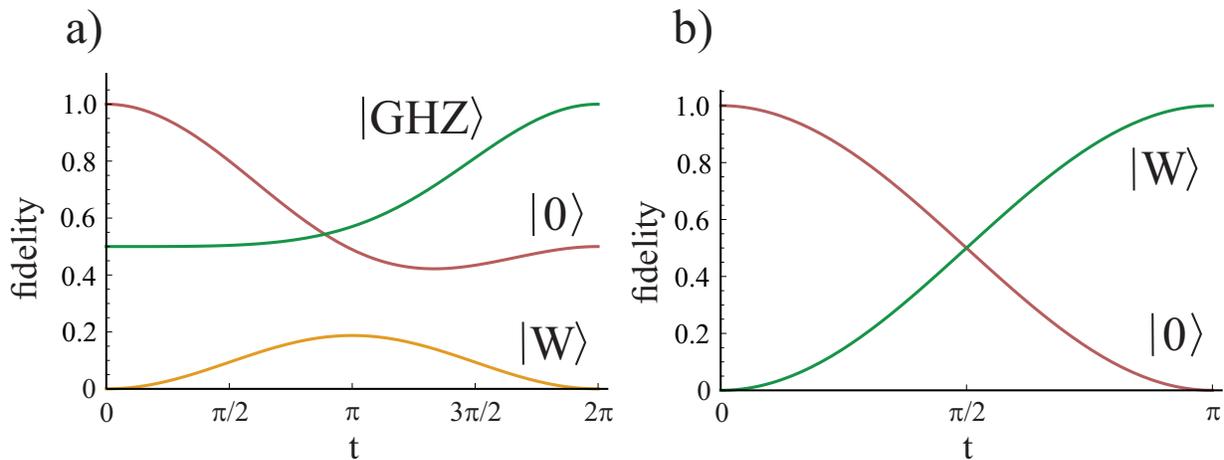}
\caption{\label{amplitudes} Change of {\color{black} fidelities} during {\color{black}the optimal} unitary evolution generated by Hamiltonians with at most two-body interactions for {\color{black}three-qubit} a) GHZ and b) W state. Each curve is associated with a contribution from the corresponding state. For the GHZ, natural interactions incorporate additional W state components and thus lead to a slowdown. The saturation of QSL for the W state is a consequence of two-dimensional evolution. }
    \end{figure}

{\color{black}The optimal} unitaries generated by two-body Hamiltonians can incorporate more states than just the initial and target in the evolution and thus result in the slowdown. In the case of three-qubit GHZ and W state, the change of {\color{black} fidelities} associated with the corresponding states in time has been depicted in Fig.~\ref{amplitudes}. As noted before, the optimal Hamiltonians, featuring a two-dimensional flow of {\color{black} fidelities} between the initial and an appropriate orthogonal state, cannot be always realised without {\color{black} many-body} terms. This statement is true for the GHZ state~\cite{Brody_2006}. The time evolution of {\color{black} fidelities} generated by at most two-body Hamiltonians for this state leads to the emergence of non-zero contributions from the W state, see Fig. \ref{amplitudes}a). Before fidelity with the target state reaches 1, the {\color{black} fidelities} associated with the additional state have to be damped. This clearly needs some resources and is the reason for the slowdown. The low dimensional behaviour of {\color{black} fidelities} in the two-body interactions regime for the W state can be observed in Fig.\ref{amplitudes}b). As before, this confirms the analytical predictions and is the reason for the saturation of the general QSL in Eq.~(\ref{MT}).

\begin{figure}[ht]
    \centering
        \includegraphics[width=0.57\linewidth]{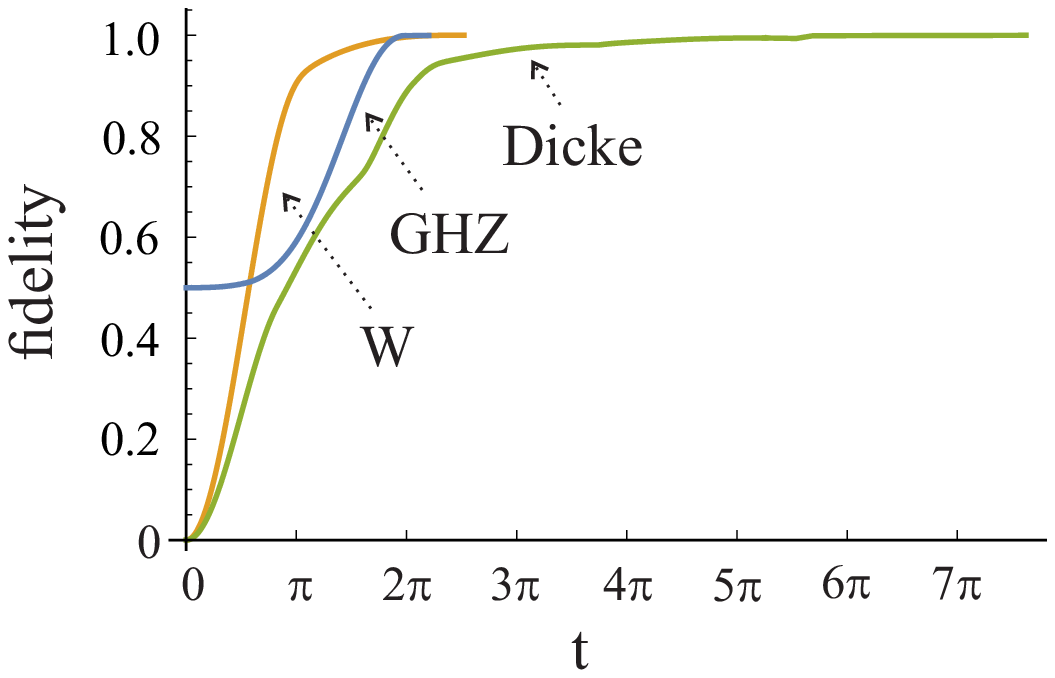}
    \caption{Fidelity optimisation for the four-qubit states. More excitations and elements in the superposition lead to the slower evolution to the Dicke $| D_4^2 \rangle$ state in the two-body regime. As opposed to the three-qubit case, it is here possible to obtain the GHZ state faster than the W state. Note that the minimal time for the GHZ state has not changed despite adding an additional qubit{\color{black}, see Fig.~\ref{three_qubits}}.}
    \label{four_qubits}
\end{figure}

In the four-qubit scenario, there is a clear difference in the dynamics of fidelity optimisation for GHZ, W and two-excited Dicke state, see Fig.~\ref{four_qubits}. {\color{black} While the quantum speed limit for the three-qubit W state was saturated with the two-body interaction } in the four-qubit case it is not and the GHZ state can be achieved in a shorter time{\color{black}, see Fig.~\ref{four_qubits}}. The minimal time difference between one- and two-excited Dicke states is significant and results from the greater amount of flips in each of the basis states and the number of elements in the superposition. The computational difficulty of optimising the fidelity to the Dicke state was considerably greater than that of the W and GHZ states. It should be noted that, in the standard quantum speed limit, the minimum time required to obtain a perfect fidelity between the $| D_4^2 \rangle$ and $\ket{0000}$ states is $\pi$, as they are orthogonal. However, we have observed that the minimal time required to achieve a perfect fidelity using two-body interactions is approximately $7.5$ times greater.

\subsection{GHZ states}
\label{sec:ghz}

\begin{figure}[ht]
    \centering
    \includegraphics[width=1\linewidth]{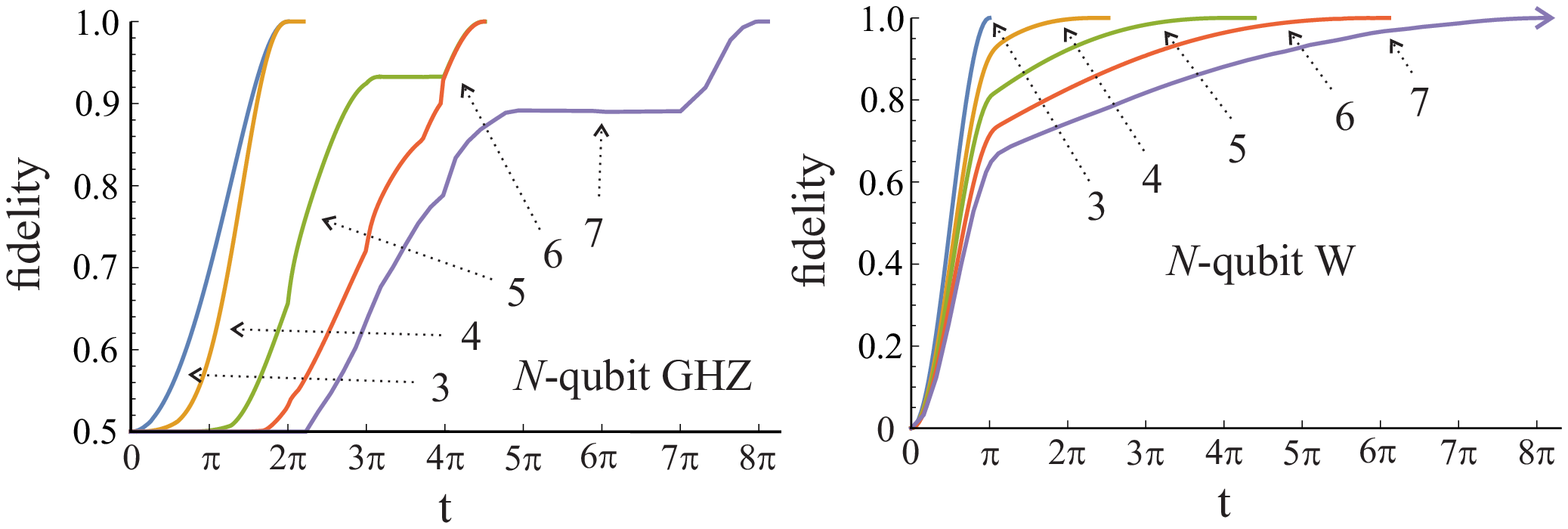}
    \caption{Maximal fidelity $\mathcal{F}$ as a function of time. The left panel presents results for $N=3,4,5,6,7$ qubits in GHZ states. The minimal time needed to obtain the unit fidelity with two-body interactions is the same for the pairs of neighbouring curves, i.e. the ones associated with $3$ and $4$ or $5$ and $6$ particles. For the smaller number of parties, its behaviour is clearly described by a single trend while for the $5,6$ and $7$ qubits case, it has to be some composition of different functions. The right panel describes $N=3,4,5,6,7$ qubits in the $W$ state obtained with two-body interactions only. {\color{black} All the curves} have a similar character. 
    Fidelity grows rapidly and smoothly
    until time $\pi$, where the three-qubit state reaches $\mathcal{F}=1$. Then the qualitative transition occurs and each curve approaches the desired value in time $t_{\min}$ that is strictly growing with the number of particles. }
    \label{speedlimits}
\end{figure}

The optimisation results obtained for $N=3,4,5,6,7$ qubits in the GHZ state are shown in Fig.~\ref{speedlimits}. An interesting effect associated with $t_{\min}$ for the neighbouring pairs of curves can be observed. The minimal time needed to obtain the GHZ state with two-body interactions only is the same for $3$ and $4$ as well as $5$ and $6$ particles. This means that the amount of time (energy) needed for the creation of strong multipartite correlations is not strictly rising with the number of qubits. 
The observed pattern for the two-body GHZ QSL is extrapolated as follows:
\begin{equation}
    t_{\min}^{\mathrm{GHZ},2}= \pi \lceil{N/2}\rceil ^2/2,
    \label{EQ_TMINGHZ}
\end{equation}
{\color{black}where $\lceil \cdot \rceil$ is a ceiling function. This agrees with the performed calculations and gives the expected behaviour of the time being grouped by pairs.}
This result can be easily comprehended in the sequential framework. For comparison, note that a product state of $N$ qubits on which one applies $\lceil{N/2}\rceil$ two-qubit and one-qubit gates is never genuinely $N$-party entangled, for $N > 3$.
The time it takes to apply all these gates under the constraint that the total energy range (of the entire sequence) is $|E| = 1$ can be estimated as follows.
Since there are $\lceil{N/2}\rceil$ gates, the energy per gate is $E_1 = 1 / \lceil{N/2}\rceil$. The corresponding maximal energy standard deviation is $\Delta H = E_1 / 2$. The total time is therefore
\begin{equation}
t_{\textrm{seq}} \ge \lceil{N/2}\rceil \, \, 2 \lceil{N/2}\rceil \pi/4,
\end{equation}
where $\lceil{N/2}\rceil \pi/2$ is the QSL bound for a single gate assuming the overlap between the input and the output of the gate is equal to $1 / \sqrt{2}$, as one would have to create a global GHZ state.
The shortest sequential time is, therefore, the same as $t_{\min}^{\mathrm{GHZ},2}$ but we emphasise again that the sequence is not capable of producing genuine $N$-party entanglement.

{\color{black}
To create a GHZ state of $N$-qubits in a sequential approach, a Hadamard gate followed by $N-1$ CNOT gates are typically used. The Hamiltonian associated with the Hadamard gate at the time $t=1$ is given by
\begin{equation}
    H=\pi \left[\begin{array}{cc}
        \frac{1}{4}(\sqrt{2}-2) & \frac{1}{2 \sqrt{2}} \\
        \frac{1}{2 \sqrt{2}} & -\frac{1}{4}(2+\sqrt{2}) \\
    \end{array}\right],
\end{equation}
which can be normalised by adding and dividing by $\pi$. This results in the desired superposition of two basis states at $t=\pi$. Similarly, the CNOT gate also has a normalised time of $t=\pi$. Hence, the minimal time required to generate the $N$-partite GHZ state after normalising the energy of the entire sequence is $t_{\min}^{seq} = N^2 \pi$. Recall that the minimum time obtained in the two-body regime, see Eq.~(\ref{EQ_TMINGHZ}), is less than or equal to $t_{\min}^{seq}$. 
}

The natural effect of energy normalisation is the delay in a significant change of the initial state under unitary evolution. In a sense, it resembles the concept of static friction in classical mechanics, see Fig. \ref{speedlimits}. Due to the high overlap between the initial and final state, this effect is clearly much more visible in the evolution to the GHZ states.

\subsection{W states}

The maximisation results for the W states of $N=3,4,5,6,7$ qubits are presented in the right panel of Fig.~\ref{speedlimits}. Analogous to the GHZ case the minimal times are growing with the number of particles but do not join in pairs. The character of the optimised functions is similar for each curve. 

The findings presented here can also be interpreted as optimal in the sense that they enable the production of the desired states in a shorter time compared to the conventional sequential approach. {\color{black} This has been already discussed in Sec.~\ref{sec:ghz} for the GHZ states. In the case of the W states, even for three qubits, the standard circuit consists of five gates, including one Hadamard and two CNOT gates discussed in the previous subsection. Recall that using our approach we have shown that for $N=3$ the QSL can be saturated for the W state. This makes our argument even stronger.} \\

\begin{figure}[ht]
    \centering
    \includegraphics[width=0.95\linewidth]{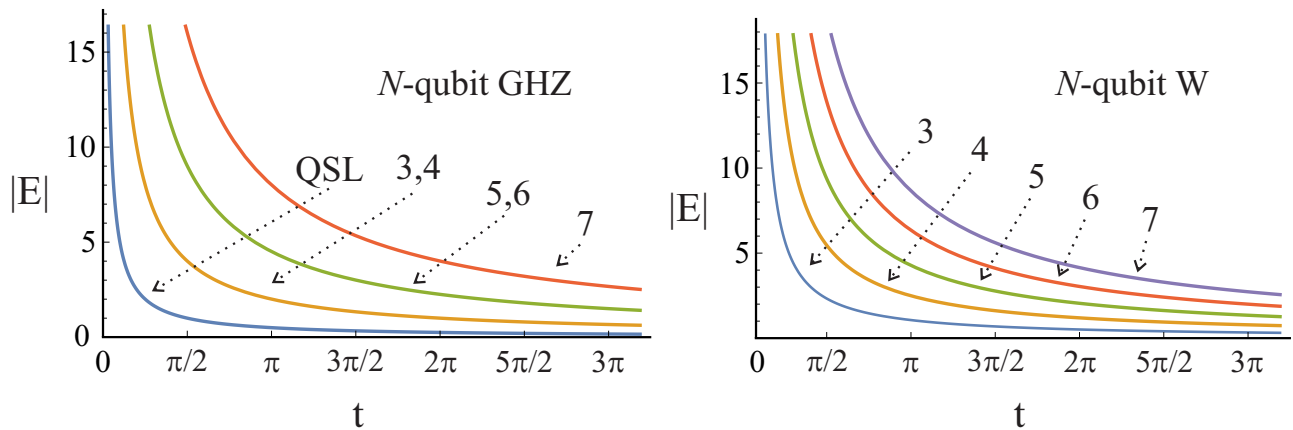}
    \caption{Required energy range for the generation of the GHZ and W state in the time $t$ with two-body interactions only. The blue curve corresponds to the standard quantum speed limit with no constraints on the nature of interactions. Note that for the W state, it coincides with $N=3$. This figure answers the following question: Having the energy range $|E|$ available, what is the minimal time in which one can generate the desired state with two-body interactions?}
    \label{trade-off}
\end{figure}

Another interesting question, in terms of resources, concerns the achievable rate of producing a multipartite entangled state with the natural interaction, given an energy range $|E|$ that is either greater or smaller than $1$. Our normalisation method allows us to extend or tighten the energy range of $H$ by simply multiplying it by $|E|$. This results in eigenvalues $E_i \in [0,|E|]$, with {\color{black} $|E|$ being inversely proportional to} time. The time-energy trade-off resulting from the presented QSLs for GHZ and W states has been presented in Fig.~\ref{trade-off}.   

\subsection{Five qubit AME state}
\label{sec:ame}
A somewhat different example is provided by the AME state {\color{black} defined in Eq.~(\ref{ame52}).} 
The isotropic and symmetric Hamiltonian evolution yields $\mathcal{F}\approx 0.1$ even for times $t \gg 1$.
This poor fidelity is a consequence of the symmetry of that state. Although it is not completely symmetric, it remains invariant under $\mathcal{P}_{24}\mathcal{P}_{35}$, where $\mathcal{P}_{ij}$ denotes a swap between the $i$-th and $j$-th particle. To achieve the AME state we, therefore, introduce a new symmetry of the Hamiltonian that aligns with the desired state's symmetry, i.e. the Hamilton operator should remain invariant under $\mathcal{P}_{24}\mathcal{P}_{35}$. 
This leads to 29 parameters that require optimisation, in contrast to the 9 parameters involved in the previous case.

Our analysis of the speed of unitary evolution for the AME state shows that there exists a long-range plateau after achieving $\mathcal{F}=0.99$. This implies that it takes considerably less time to evolve the state from $0-99\%$ fidelity than to enhance it by only $1\%$. Based on this observation and the number of parameters that require optimization, we calculated the time for the fidelity to surpass $99\%$. The final result is $t_{99\%}=10.72$, as presented in Table.~\ref{res}.

\subsection{Generating Hamiltonians}

The numerical optimisation has allowed us to derive the analytical expressions for the Hamiltonians that generate the fastest evolution to the $\mathrm{W}$ and $\mathrm{GHZ}$ states. The explicit formulas, including normalisation factors, can be found in~\ref{sec:appII}. We shall use the conventional periodic boundary conditions $\sigma^{(N+1)}_{\mu}=\sigma^{(1)}_{\mu}$. Based on our results, the optimal (unnormalised) Hamiltonian for the $W$ states takes on a simple form
\begin{equation}
   H= \sum_{i=1}^N \sum_{j=i+1}^N \sigma_x^{(i)}  \sigma_z^{(j)} + \sigma_z^{(i)}   \sigma_x^{(j)} -(N-3)\sigma_x^{(i)}.
\end{equation}

For the $\mathrm{GHZ}$ state with an odd number of qubits, the general expression for the fastest two-body Hamiltonian is:
\begin{equation}
H= \sum_{i=1}^N \sum_{j=i+1}^N  \sigma_x^{(i)}   \sigma_x^{(j)} -\sigma_y^{(i)}   \sigma_y^{(j)} + \sigma_z^{(i)}   \sigma_z^{(j)} + 2 \sigma_y^{(i)}.
\end{equation}
Surprisingly, this Hamiltonian is realised by a spin-$1/2$ chain beyond the nearest-neighbour $XXY$ coupling and magnetic field in the $y$ direction.

Results for the GHZ state with an even number of qubits are presented in ~\ref{sec:appII}. As an example, the Hamiltonian for $N=4$ is
\begin{equation}
H=\sum_{i=1}^4  \sum_{j=i+1}^4 \frac{1}{2} (\sigma_x^{(i)}   \sigma_y^{(j)}+ \sigma_y^{(i)}   \sigma_x^{(j)} ) +  \sigma_y^{(i)}   \sigma_y^{(j)}  + \left(\frac{1}{2} -\frac{1}{\sqrt{2}}\right) \sigma_z^{(i)}   \sigma_z^{(j)}.
\end{equation}

\section{Conclusion}

We considered the quantum speed limit for the generation of multipartite entangled states in the presence of physically motivated constraints. In particular, a natural constraint in multipartite systems is that the Hamiltonian contains at most two-body interaction terms. 
We showed that such Hamiltonians allow one to generate genuinely multipartite entangled states, such as W, GHZ, Dicke {\color{black}(up to fidelity 1)} and AME$(5,2)$ {\color{black}(up to fidelity 0.99)}, from an initial product state. Moreover, we have established the fastest times to generate them,
taking into account the limitations imposed by the energy and type of interaction. 
The optimal Hamiltonians were also derived. We have found that the amount of time (energy) required for the creation of strong multipartite correlations does not have to rise strictly with the number of qubits. This phenomenon has been observed in the context of GHZ states, which are generated in the same minimal time for $N=2,3$ and also for $N=4,5$. We conjecture that this trend continues and we extrapolated the shortest time in a closed formula for any $N$.
Our results demonstrate that in terms of time resources, our approach outperforms the sequential model. This sets the direction for fast entanglement generation with a lower noise level in specific tasks. Future research concerning this problem could concentrate on the analytical derivation of the suggested quantum speed limits and a more detailed study of the generating Hamiltonian implementation.

\section*{Acknowledgements}

PC acknowledges the support of the Polish National Science Centre (NCN) within the Preludium Bis project (Grant No. 2021/43/O/ST2/02679). PK is supported by the Polish National Science Centre (NCN) under the Maestro Grant No. 2019/34/A/ST2/00081. WL acknowledges support from the Foundation for Polish Science (IRAP project ICTQT, Contract No. 2018/MAB/5, co-financed by EU via Smart Growth Operational Programme).


\appendix

\section{Eigenvalues of 3 qubit system with 2-body interactions}
\label{sec:appI}

\subsection{Optimal creation of W state}
\label{sec:appI.1}

Let us start with the Hamiltonian {\color{black} without} one-body terms
\begin{equation}
H=\sum_{i=1}^N \sum_{j=i+1}^N \sum_{\mu_i,\mu_j} h_{\mu_i \mu_j}\sigma^{(i)}_{\mu_i}\sigma^{(j)}_{\mu_j}, \quad h_{\mu_i \nu_j}=h_{\mu_k \nu_l}=h_{\nu_k \mu_l}.
\label{APP_H}
\end{equation}
The eigenvalues of the unnormalised isotropic and symmetric {\color{black} complete interaction graph} Hamiltonian are 
\begin{equation}
    E= \left[
    \begin{array}{l}
        -h_{ xx}-h_{yy}-h_{zz} \\
        -h_{ xx}-h_{yy}-h_{zz} \\
        -h_{ xx}-h_{yy}-h_{zz}\\
        -h_{ xx}-h_{yy}-h_{zz}\\
        h_{ xx}+h_{yy}+h_{zz} -2 F \\
        h_{ xx}+h_{yy}+h_{zz} -2 F \\
        h_{ xx}+h_{yy}+h_{zz} +2 F \\
        h_{ xx}+h_{yy}+h_{zz} +2 F \\
    \end{array}\right],
    \label{zerobeig}
\end{equation}
where $F= [h_{ xx}^2 +h_{yy}^2+h_{zz}^2+3(h_{xy}^2+h_{xz}^2+h_{yz}^2)-h_{ xx}h_{yy}-h_{ xx}h_{zz}-h_{yy}h_{zz}]^{1/2}$. Such Hamiltonian has at least two energy levels. 
Two unnormalised eigenvalues occur for
$$h_{ xx}=\frac{-h_{yy}h_{zz}+h_{xy}^2+h_{yz}^2+ h_{xz}^2}{h_{yy}+h_{zz}},$$
and are given by $E=\lbrace -\eta,-\eta,-\eta,-\eta,-\eta,-\eta,3\eta,3\eta,3\eta \rbrace$, where $$\eta=(h_{yy}^2+h_{zz}^2+h_{xy}^2+h_{yz}^2+h_{xz}^2+h_{yy}h_{zz})/(h_{yy}+h_{zz}).$$ This choice of $h_{\mu\nu}$ preserves the hermiticity of $H$ and allows the creation of the W state via the unitary evolution in the optimal time. One verifies that in this case $\Delta H = 1/2$ as reported in the main text.

\subsection{Optimal creation of GHZ state}
\label{sec:appI.2}

{\color{black} A general Hamiltonian (\ref{APP_H})} cannot produce the GHZ state via the unitary evolution due to vanishing overlap $\bra{111}U\ket{000}$. Accordingly, we need at least one non-zero $b_{\mu}$ coefficient. Without loss of generality, we assume $b_y \neq 0$. The modified eigenvalues are
\begin{equation}
    E=\left[
    \begin{array}{l}
        -h_{ xx}-h_{yy}-h_{zz} +b_y \\
        -h_{ xx}-h_{yy}-h_{zz} +b_y \\
        -h_{ xx}-h_{yy}-h_{zz} - b_y \\
        -h_{ xx}-h_{yy}-h_{zz} - b_y \\
        h_{ xx}+h_{yy}+h_{zz} -\frac{1}{2} \sqrt{V}-\frac{1}{2} \sqrt{\tilde{V}} \\
        h_{ xx}+h_{yy}+h_{zz} -\frac{1}{2} \sqrt{V} +\frac{1}{2} \sqrt{\tilde{V}}\\
        h_{ xx}+h_{yy}+h_{zz} +\frac{1}{2} \sqrt{V} -\frac{1}{2} \sqrt{\tilde{V}}\\
        h_{ xx}+h_{yy}+h_{zz} +\frac{1}{2} \sqrt{V} +\frac{1}{2} \sqrt{\tilde{V}}\\
    \end{array}\right],
\end{equation}
where $V$ and $\tilde{V}$ are complicated functions of all the parameters. The only possibility for an additional degeneracy, other than the one occurring for the first two pairs of energies, is when $\sqrt{V}=\sqrt{\tilde{V}}$. This gives at least 5 different energy levels which can be arranged for the energy standard deviation in the initial state equal to $\Delta H = 1/8$.

\section{Explicit formulas for the generating Hamiltonians}
\label{sec:appII}

\subsection{Optimal Hamiltonians for W state}

The following Hamiltonians $H_N$ lead to the fastest evolution to the $N$-qubit W state in the  complete graph configuration: 

\begin{eqnarray}
    H_3&=&\frac{1}{4\sqrt{3}}\left[\sum_{i=1}^3 \sum_{j=i+1}^3\sigma_x^{(i)}  \sigma_z^{(j)} + \sigma_z^{(i)}   \sigma_x^{(j)} + 2 \sqrt{3} \openone \right] \\
    H_4&=&\frac{1}{4\sqrt{22}}\left[\sum_{i=1}^4 \sum_{j=i+1}^4 \sigma_x^{(i)}  \sigma_z^{(j)} + \sigma_z^{(i)}   \sigma_x^{(j)} -\sigma_x^{(i)} + 2 \sqrt{22} \openone \right] \\
    H_5&=&\frac{1}{36}\left[\sum_{i=1}^5 \sum_{j=i+1}^5 \sigma_x^{(i)}  \sigma_z^{(j)} + \sigma_z^{(i)}   \sigma_x^{(j)} -2\sigma_x^{(i)} + 18 \openone \right] \\
    H_6&=&\frac{1}{4\sqrt{3(41+\sqrt{921})}}\left[\sum_{i=1}^6 \sum_{j=i+1}^6 \sigma_x^{(i)}  \sigma_z^{(j)} + \sigma_z^{(i)}   \sigma_x^{(j)} -3 \sigma_x^{(i)} \right.\\
    &-& \left. 2\sqrt{3(41+\sqrt{921})}\openone \right] \nonumber \\
    H_7&=&\frac{1}{32(16+\sqrt{31})}\left[\sum_{i=1}^7 \sum_{j=i+1}^7 \sigma_x^{(i)}  \sigma_z^{(j)} + \sigma_z^{(i)}   \sigma_x^{(j)} -4\sigma_x^{(i)} \right.\\
    &-& \left. 2 (16+\sqrt{31}) \openone \right] \nonumber
\end{eqnarray}

\subsection{Optimal Hamiltonians for GHZ state}

Below are the explicit formulas for the Hamiltonians $H_N$ generating the fastest evolution to the $N$-qubit GHZ states in the two-body interaction regime. All formulas, except the first one, pertain to the complete interaction graph: 

\begin{eqnarray} 
H_3^{\mathcal{I}=1} &=& \frac{1}{10}\left[\sum_{i=1,2} \left( \sqrt{2} \left(\sigma_x^{(i)}   \sigma_x^{(i+1)} +
\sigma_z^{(i)}   \sigma_z^{(i+1)} \right)+  \sigma_y^{(i)} \right)  + 5 \openone\right]\\
H_3&=& \frac{1}{16}  \left[ \sum_{i=1}^3 \sum_{j=i+1}^3 \left( \sigma_x^{(i)}   \sigma_x^{(j)} - \sigma_y^{(i)}   \sigma_y^{(j)}  + \sigma_z^{(i)}   \sigma_z^{(j)} + 2 \sigma_y^{(i)} \right)+ 9 \openone \right]\\
H_4 &=&\frac{1}{8 \sqrt{2}} \left[ \sum_{i=1}^4  \sum_{j=i+1}^4 \left(\frac{1}{2} (\sigma_x^{(i)}   \sigma_y^{(j)}+ \sigma_y^{(i)}   \sigma_x^{(j)} ) +  \sigma_y^{(i)}   \sigma_y^{(j)} \right. \right.\\ 
  &+& \left. \left. \left(\frac{1}{2} -\frac{1}{\sqrt{2}}\right) \sigma_z^{(i)}   \sigma_z^{(j)}    \right) + \openone (3-5 \sqrt{2})\right] \nonumber \\
H_5&=& \frac{1}{36}\left[\sum_{i=1}^5 \sum_{j=i+1}^5 (\sigma_x^{(i)}   \sigma_x^{(j)} -\sigma_y^{(i)}   \sigma_y^{(j)} + \sigma_z^{(i)}   \sigma_z^{(j)} + 2 \sigma_y^{(i)} ) + 20 \openone \right]\\
H_6&=& \frac{1}{36}\left[\sum_{i=1}^6 \sum_{j=i+1}^6 (-(\sigma_x^{(i)}   \sigma_y^{(j)} + \sigma_y^{(i)}   \sigma_x^{(j)}) - \sigma_z^{(i)}   \sigma_z^{(j)} ) + 21 \openone \right] \\
H_7&=& \frac{1}{64}\left[\sum_{i=1}^7 \sum_{j=i+1}^7 (\sigma_x^{(i)}   \sigma_x^{(j)} -\sigma_y^{(i)}   \sigma_y^{(j)} + \sigma_z^{(i)}   \sigma_z^{(j)} + 2 \sigma_y^{(i)} ) + 35 \openone \right]
\end{eqnarray}

\providecommand{\newblock}{}

\end{document}